\documentclass[12pt]{iopart}

\usepackage{graphicx}
\usepackage{color}
\begin{document}
\title{Gyrokinetic resonant theory of low frequency electromagnetic perturbation }

\author{Shuangxi Zhang}

\address{ Institute of Fluid Physics, China Academy of Engineering Physics, Mianyang 621999, China \\
}
\ead{zshuangxi@gmail.com}

\date{\today}

\begin{abstract}
It's pointed out that if the normalized amplitude of low frequency electromagnetic perturbation is larger than the characteristic small parameter which is the ratio of gyro period over transiting period, and if resonance happens between $\omega$ and $\mathbf{k}\cdot \mathbf{v}$, modern gyrokinetic theory violates the basic property of near identity transformation, which is supposed to be obeyed by Lie perturbed transformation theory. A modification is given to overcome this problem by not requiring  all components in the first order Lagrangian 1-form equaling zero. A numerical example is given as an application of the new theory.
\end{abstract}

\maketitle

\section{INTRODUCTION}\label{sec1}

Modern gyrokinetic theory (GT) is a strong theoretical tool for numerical calculating of the orbit of charged particles immersed in strong magnetic field, since the fast gyro angle is decoupled from the dynamic equations of other degrees of freedom in the new coordinate system \cite{1983cary,1990brizard,2007brizard1,2009cary,wwleejcp1987,xuxueq1991,zhlinpop1995,Idomuranf2009,
1999honqingpop,biancalanipop2016}. The whole scheme of modern GT is to apply Lie perturbed transformation theory (LPTT) to non-canonical guiding center Lagrangian 1-form to find a new coordinate frame to recover magnetic moment as an adiabatic invariant by getting rid of $\theta$ dynamics \cite{1983cary,1990brizard,1988hahm}. The basic property of LPTT is that it's a near identity transformation (NIT)\cite{1983cary}.

It's pointed out in Ref.(\cite{2016nfshuangxi}) that the application of the resonant perturbed theory given by John Cary \cite{1983cary} to high frequency circular polarized wave driving charged particle in strong magnetic field in Ref.(\cite{gunyoungpop2007}), violates the property of NIT in some range of perturbed amplitude. In this paper, we found that if the scheme of modern GT is carried out, even the low frequency electromagnetic perturbation could also cause Lie perturbed coordinate transformation to violate NIT when resonance happens between $\omega$ and $\mathbf{k}\cdot \mathbf{v}$, only if the normalized amplitude of perturbation is large than the characteristic small parameter $\varepsilon$, which equals the ratio between gyro period and transiting period. Here, $i$ represents the $ith$ coordinate,  $\omega$ is the frequency of wave, $\mathbf{k}$ is the wave vector and $\mathbf{v}$ is the velocity of gyrocenter. The low frequency here means that the frequency is much lower than the gyro frequency of relevant charged particle.

The basic reason can be traced back to the dealing method for resonant branches in Modern GT. Modern GT requires ${\Gamma}_{1i}=0$ for each $i$ except $i\ne 0$ and ${\Gamma}_1$ is the first order 1-form in new coordinates. This requirement introduces an almost constant term originating from the resonant perturbation, to the differential equation of $S_1$ and induces the secularity property to $S_1$. To avoid the secularity of $S_1$, the usual way is to move this resonant branches out of the equation of $S_1$, but add it into the first order energy $H_1$, which is defined to be $-\Gamma_{10}$. However, the relevant generators included by the formula of coordinate transformation still includes those resonant branches.

The method to overcome this problems is to keep in the relevant original ${\Gamma}_{1i}$s the resonant branches, which will not appear in the equations for $S_1$ and relevant generators, rather than requiring all ${\Gamma}_{1i}=0$ except $i\ne 0$ as the modern GT does. In the real physical environment, for each low frequency wave of $\omega$ and $\mathbf{k}$, there inevitably exists particles whose gyrocenter velocity can cause resonance with the wave, since the velocity of particle ensemble has a very broad distribution. Therefore, modern GT inevitably violates NIT for any low frequency electromagnetic perturbation, which makes sure our modified theory more reasonable.

Besides, it's found that our modified theory is almost the same as that named 'Symplectic Representation' by Brizard in his doctor dissertation \cite{1990brizard}, where he gave two ways to carry out the LPTT for the perturbation in guiding center system. The other one is named as 'Hamiltonian Representation', which is adopted by the following researchers and called modern GT in this paper. But Brizard in that paper didn't discuss resonant behavior, neither did he study that for what kind of problems, which one of the two methods is preferred. In this paper, our discovering claims that the 'Symplectic Representation' rather than 'Hamiltonian representation' should be adopted to carry out the LPTT with low frequency electromagnetic perturbation.

The arrangement of this paper is as follows.
In Sec.~\ref{sec2}, the violation of NIT is presented.
In Sec.~\ref{sec3}, the scheme for preventing the violation of NIT is given, and the comparison between the old and new theory is carried out.
In Sec.~\ref{sec4}, a simple example is given as an application of the new theory.

\section{The violation of NIT of modern GT with resonant electromagnetic perturbation. }\label{sec2}

In this paper, for the convenience of notation, the original guiding center plus the time is chosen to be $\mathbf{\bar Z} = ({\bf{\bar X}},\bar {U},\bar {\mu} ,\bar{ \theta} ,t)$, while the gyrocenter plus the time is chosen as ${\bf{Z}} = ({\bf{X}},U,\mu ,\theta ,t)$. And the coordinate transformation formula for LPTT is ${\bf{\bar Z}} = {e^{{G^i}{\partial _{{Z^i}}}}}{\bf{Z}}$ where $G^i$ is the infinitesimal generator of each $Z^i$, and $G^0=0$ is assumed. Here, superscript $0$ represents the time. In this paper, $\bar{f}$ means function of $({\bf{\bar X}},\bar {U},\bar {\mu} ,\bar{ \theta} )$, unless other arguments is obviously given. The details of the scheme of modern GT with electromagnetic perturbation is given in appendix Sec.({\ref{sec11}}). The operator ${G}^i{\partial _{{{Z}^i}}}$ is dimensionless. NIT requires the value of $|G^i|/Z_0^i$  to  be much smaller than one, where $Z^i_0$ is used to normalize each $Z^i$.

The Fourier analysis of $\mathbf{A}_1$ can be expressed as
\begin{equation}\label{e1}
{{\bf{A}}_1}\left( {{\bf{X}},t} \right) = \sum\limits_{{\bf{k}}'} {\left( {{A_{c{\bf{k}}'}}\cos \left( {{\bf{X}}\cdot{\bf{k}}' - {\omega _{k'}}t} \right) + {A_{s{\bf{k}}'}}\sin \left( {{\bf{X}}\cdot{\bf{k}}' - {\omega _{k'}}t} \right)\;} \right)\;\;\;}
\end{equation}
Here, we assume that fourier branch $(\omega_k,\mathbf{k})$ satisfies the resonant condition ${\omega _k} - {\bf{v}}\cdot{\bf{k}} \approx 0$ in a resonant layer where the gyrocenter velocity is $\mathbf{v}$, and only the cosine branch exists. According to the operation of gyroangle averaging and the removing of secularity from gauge function $S$ in appendix, this branch is removed from the gauge function $S_1$. We noticed that in $G^U$ in Eq.(\ref{a2}) this resonant branch is left to the first term on the right hand side. If other non-resonant Fourier branches are ignored except the resonant one, the dimensionless value of $G^U$ after normalization by $v_t$ can be reformulated to be
\begin{equation}\label{e2}
{G^U} \approx \varepsilon^{-1} {A_{ck}}\cos ({\bf{X}}\cdot{\bf{k}} - {\omega _k}t),
\end{equation}
where we made the normalization ${G^U} \to {G^U}/{v_t},{A_{ck}} \to {A_{ck}}/{A_0}$, and $\varepsilon \equiv  {m_i}{v_t}/e{A_0}$.
$v_t$ is the thermal velocity of ion. In SI system, the amplitude of equilibrium magnetic vector potential $A_0$ can be adopted as $1T/m$ and $v_t=10^4 m/s$, thus, $\left| {{\varepsilon }} \right| \approx {10^{-4}}$ for ions. In the resonant region, $\left| {\cos ({\bf{X}}\cdot{\bf{k}} - {\omega _k}t)} \right|$ is almost a constant. If $A_{ck}\ge 10^{-4}$ is satisfied, $|G^{U}| \ge 1$ may stand and violates the inequality $|G^{U}|\ll 1$ which should be obeyed by NIT.


\section{The scheme to avoid the violation of NIT}\label{sec3}

According to the analysis in Sec.(\ref{sec2}) and in appendix Sec.(\ref{sec11}), the violation of NIT involves term $e\bar{\mathbf{A}}_1\cdot d\bar{\mathbf{X}}$ in $\bar{\gamma}_1$ when solving $G^i$ and $S_1$ by requiring $\Gamma_{1i}=0$. The problem can be overcome by keeping this term in $\Gamma_{1\mathbf{X}}$ after carrying out the LPTT over the first order 1-form, just as 'Symplectic Representation' given in Brizard's doctor thesis does. The details is given in appendix Sec.(\ref{sec12}). The resonant branch is removed from the generators. The terms left in $G^{U}$ and $G^{\theta}$ are non-resonant terms. As before, the perturbed potential is assumed to include the resonant branch ${{\bf{A}}_1}\left( {{\bf{X}},t} \right) = {A_{ck}}\cos ({\bf{X}}\cdot{\bf{k}} - {\omega _k}t){\bf{b}}$. The contribution to the acceleration of $U$ by the resonant branch in Eq.(\ref{g11}) derived from modern GT is
\begin{equation}\label{m1}
\dot {U}_{old} =  - \left( {\frac{{eU{{\bf{B}}^*}}}{{mB_\parallel ^*}}\cdot{\bf{k}}} \right)\sin \left( {{\bf{X}}\cdot{\bf{k}} - {\omega _k}t} \right),
\end{equation}
while the contribution to the acceleration of $U$ by the resonant branch in Eq.(\ref{g24}) derived from new GT is
\begin{equation}\label{m2}
\dot {U}_{new} = \frac{e}{m}{\omega _k}\sin \left( {{\bf{X}}\cdot{\bf{k}} - {\omega _k}t} \right).
\end{equation}
When resonance happens, $\omega_k  = {\mathbf{v}}\cdot{\mathbf{k}}$ holds and $\sin \left( {{\bf{X}}\cdot{\bf{k}} - {\omega _k}t} \right)$ is almost an constant. The obvious difference between Eq.(\ref{m1}) and Eq.(\ref{m2}) is the minus sign on the left hand of Eq.(\ref{m1}), which proves that the dynamic equation of parallel velocity derived from modern GT is not right. Compared with Eq.(\ref{g11}) derived from the modern GT, Eq.(\ref{g24}) is intuitional plausible since the induced electric field appears as the driven force. The induced electric field is ${\bf{E}} =  - \frac{{\partial {\bf{A}}}}{{\partial t}}$.

However, the difference between Eq.(\ref{m1}) and (\ref{m2}) in the past simulation is hard to be observed, since for a ensemble of resonant particles, the initial phase of $\sin \left( {{\bf{X}}\cdot{\bf{k}} - {\omega _k}t} \right)$ covers the range of $(0,\pi)$ which cancels the effect of  the symbol difference between Eq.(\ref{m1}) and (\ref{m2}).

\section{a numerical application}\label{sec4}

In this paper, the numerical application of our theory is based on the simple toroidal magnetic configuration, which is
\begin{equation}\label{g28}
{\bf{B}} = \frac{{{B_0}}}{{1 + r\cos \phi /{R_0}}}\left( {{\mathbf{e}_\xi } + \frac{r}{{q{R_0}}}{\mathbf{e}_\phi }} \right),
\end{equation}
where $R_0$ is the major radius at the magnetic center and $q$ is the safety factor, and the toroidal geometry coordinate is $(r,\phi,\xi)$.  In our numerical example, the parameters are chosen as $R_0=4a$,$q=2$,$B_0=1$T, where $a$ is small radius.

To describe induced electric field driving accelerating velocity in a simple picture, the model of electromagnetic perturbation is chosen to be an magnetic potential vector of a single cosine Fourier branch, parallel to the unit vector of equilibrium magnetic field. In toroidal geometry, its expression is
\begin{equation}\label{g29}
{{\bf{A}}_1}\left( {{\bf{X}},t} \right) = {\bf{b}}A_1(r)\cos \left( { \omega t + k\phi  - n\xi } \right)
\end{equation}
where $k$ and $n$ are poloidal and toroidal wave number, and $A_1(r)$ is the amplitude of wave at radial position $r$. The resonant condition is
\begin{equation}\label{g30}
\omega  + k{\omega _\xi } - n{\omega _\phi } = 0,
\end{equation}
where ${\omega _\xi } = \frac{{d\xi }}{{dt}}$ and ${\omega _\phi } = \frac{{d\phi }}{{dt}}$. In our example, $k=1,n=1$ is chosen. The kinetic equation of $U$ given in Eq.(\ref{g24}) becomes
\begin{equation}\label{g31}
\dot U =  - \frac{{{{\bf{B}}^*}}}{{mB_\parallel ^*}}\cdot\nabla \left( {\mu B} \right) + \frac{e}{m}{A_1}(r)\omega \sin (\omega t + \theta  - \phi )
\end{equation}

The normalization quantities are $r_0=a$, $t_0=a/U_0$ , $U_0=v_{th}$, $B_0=1$T, $A_0=B_0*a$. In our example, $\omega=2\times 10^{-2}/\Omega_i $ and $A_1(r)=2\times 10^{-7}/A_0$ where $\Omega_i$ is the gyro frequency of ion based on normalization quantities.
The initial position of the charged particle is at $(x,y,z)=(4.5,0,0)$ after normalization, where the rectangular coordinates are adopted.  The initial perpendicular velocity is $v_{\perp 0}=2$. With the given equilibrium and perturbed magnetic field, and other initial conditions, the resonant parallel velocity around the initial position is about $0.2$ solved from the resonant condition Eq.(\ref{g30}). 
The normalized numerical time step is chosen to be $dt=10^{-3}$ which indicates five discrete times for one period of the wave.

\begin{figure}[htbp]
\centering
\includegraphics[height=8cm,width=8cm]{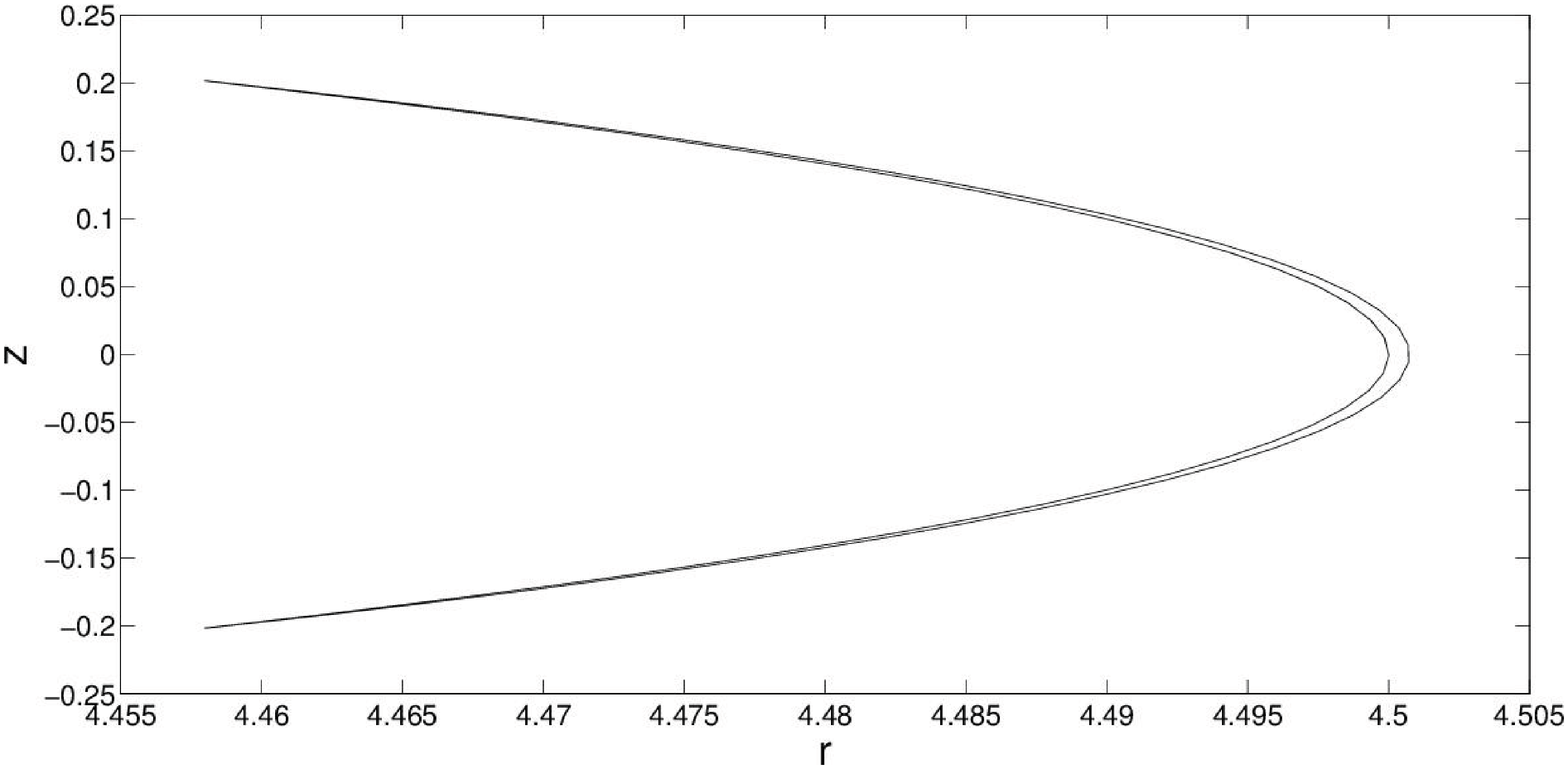}\centering
\caption{\label{gcorbit13}  The guiding center orbit of the particle without the electromagnetic perturbation with time step $dt=1e-3$. With given the initial conditions in the context, the orbit is a trapped one.}
\end{figure}

\begin{figure}[htbp]
\centering
\includegraphics[height=8cm,width=8cm]{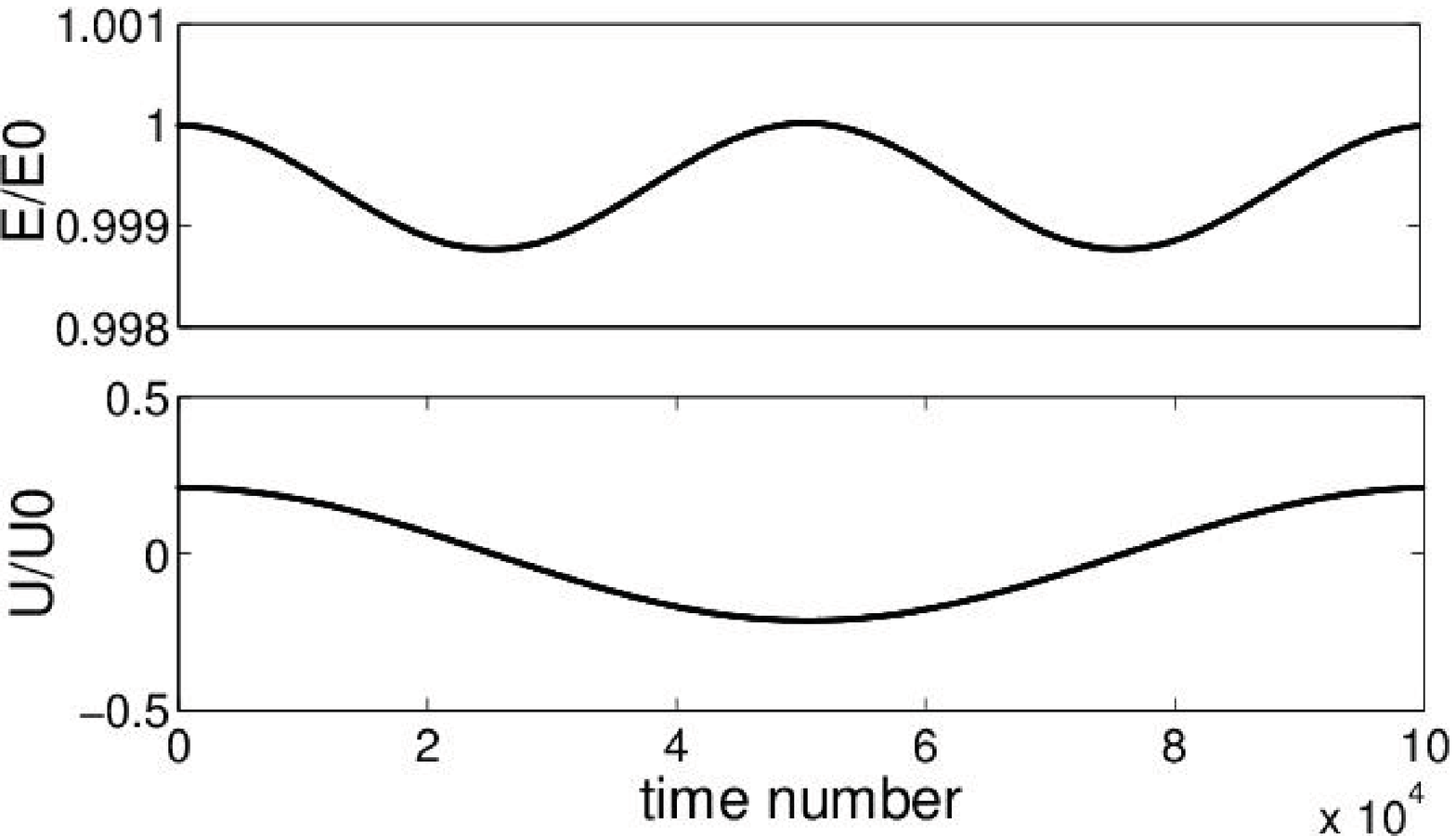}\centering
\caption{\label{gcue13} The normalized energy and parallel velocity as functions of time number without electromagnetic perturbation with time step $dt=1e-3$.}
\end{figure}

\begin{figure}[htbp]
\centering
\includegraphics[height=8cm,width=8cm]{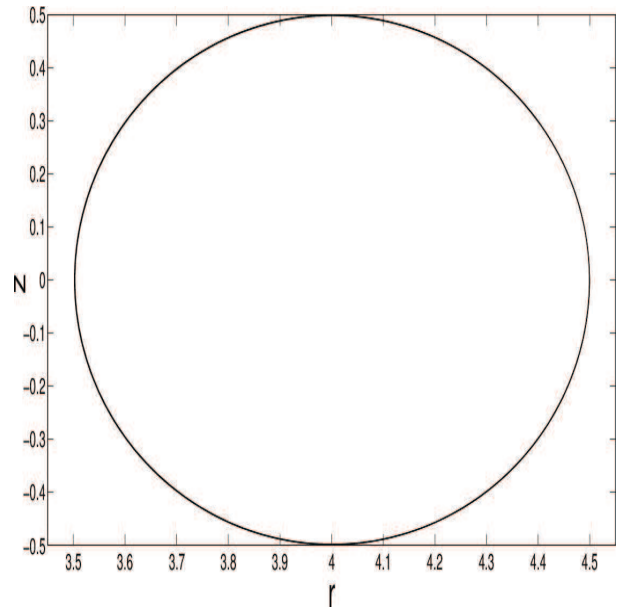}\centering
\caption{\label{mporbit13} The gyrocenter orbit of the particle with the electromagnetic perturbation and the same initial conditions is given with $dt=1e-3$. The orbit is changed from trapped one to passing one.}
\end{figure}

\begin{figure}[htbp]
\centering
\includegraphics[height=8cm,width=8cm]{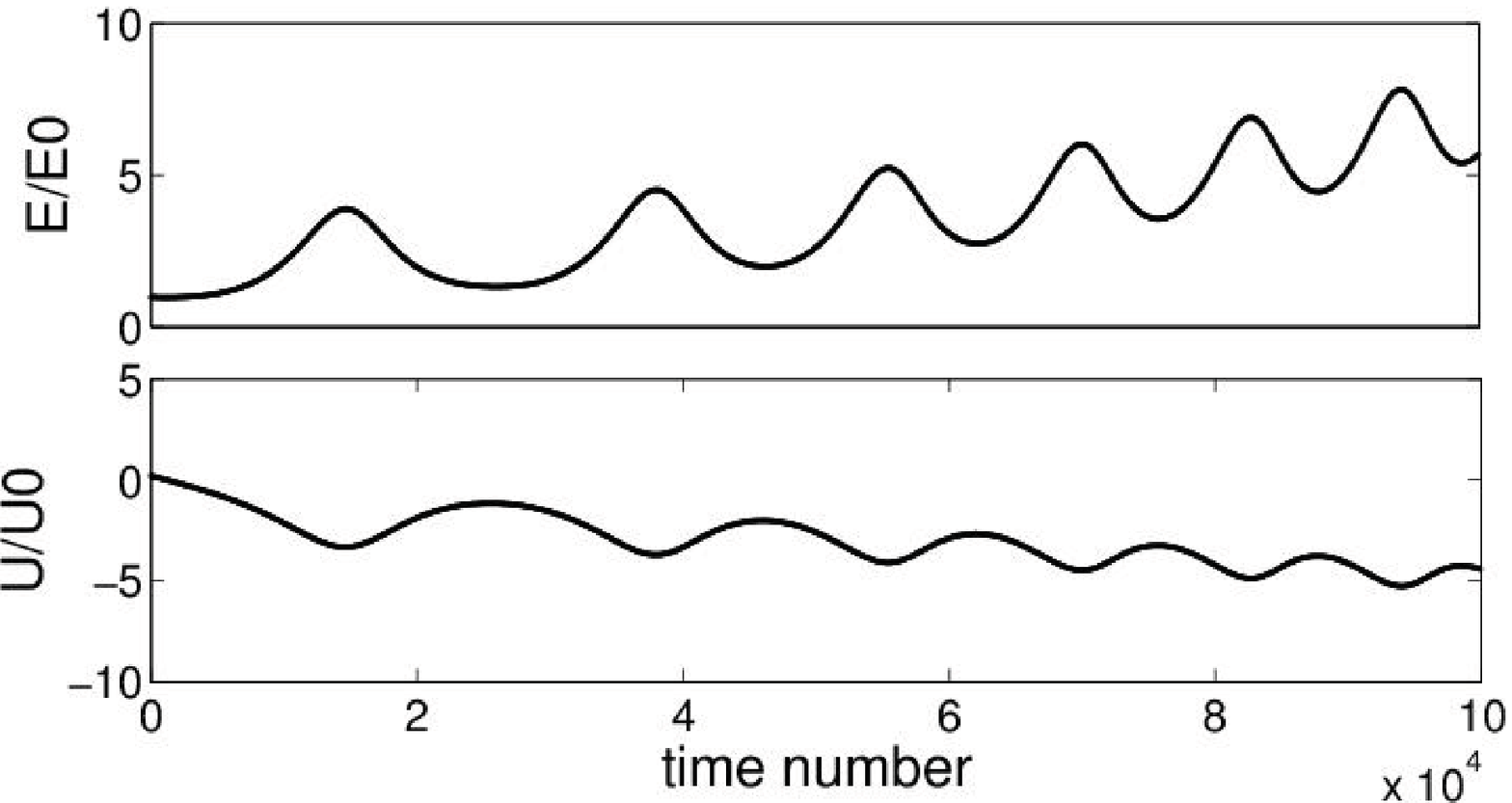}\centering
\caption{\label{mpue13} The normalized energy and parallel velocity as functions of time with electromagnetic perturbation with $dt=1e-3$. }
\end{figure}

The fourth order Range-Kutta scheme is adopted in this numerical example. Fig.(\ref{gcorbit13}) shows the trapped orbit of the guiding center of the particle with given initial conditions and equilibrium magnetic field without perturbation. The time step is $dt=10^{-3}$. The variation of normalized energy and parallel velocity of the gyrocenter along with time is given in Fig.(\ref{gcue13}). When the electromagnetic field is switched on with the same initial conditions, the orbit of the particle changes from the trapped one to the passing one as shown in Fig.(\ref{mporbit13}) with $dt=10^{-3}$. The design of the initial conditions makes the resonance happen at the beginning which can be observed by comparing Fig.(\ref{gcue13}) and Fig.(\ref{mpue13}), thus the phase of $\sin (\omega t + \theta  - \phi )$ changes slowly at the first half period during which the resonant perturbation decelerates parallel velocity to zero, then accelerate it to a large value in the opposite direction.

It's obvious in Fig.(\ref{mpue13}) that in the first half period during which resonance happens, the parallel velocity changes most and the energy transferred to the particle by the wave is much more compared with other periods. It's observed from Fig.(\ref{mpue13}) that when time goes on, the averaged parallel velocity over one period is increased although the increasing rate decreases along with time, thus a induced parallel electric field may drive charged particles to energetic ones.

\begin{figure}[htbp]
\centering
\includegraphics[height=8cm,width=8cm]{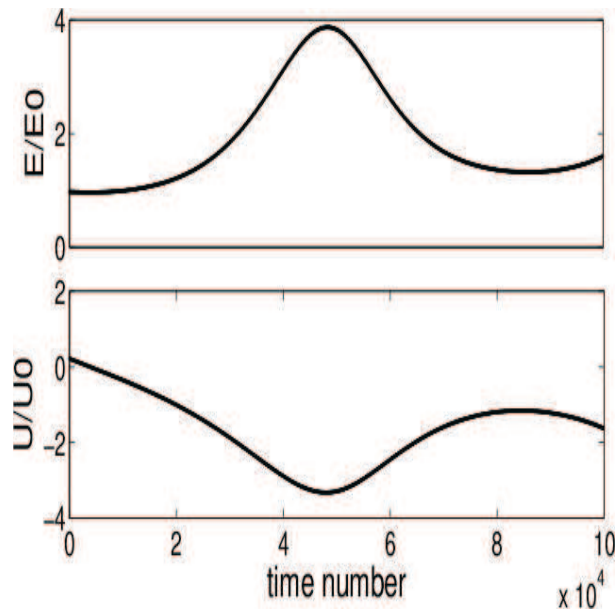}\centering
\caption{\label{mpue34} As a comparison, the first period with more accurate time step of $dt=3e-4$ is given. It's found that no obvious difference exists between this fig and Fig.(\ref{mpue13}) during the first period. }
\end{figure}

A more accurate time step $dt=3\times 10^{-4}$ is adopted in Fig.(\ref{mpue34}) to verify the numerical correctness of time step $dt=10^{-3}$. The numerical correctness of Fig.(\ref{mporbit13}) and Fig.(\ref{mpue13}) is verified by Fig.(\ref{mpue34}) based on the fact that in the first period the normalized $U$ and energy in Fig.(\ref{mpue13}) as functions of time  are almost the same with those in Fig.(\ref{mpue34}).

\section{Summary and Discussion}\label{sec9}

In this paper we pointed out that modern GT would violate NIT  with resonant electromagnetic perturbation, since $G^{U}$ can be much larger than one in some range of perturbed parameters. A modified method is given to remove this violation. Taking into account of a broad distribution of velocity of particle ensemble, the resonant behavior is inevitable. Therefore, our method is more plausible as a gyrokinetic theory dealing with the interaction between electromagnetic wave and charged particle in strong magnetic field. In fact, the error of modern GT can be inferred in an intuitive way that the kinetic equation of $U$ of Eq.(\ref{g11}) doesn't contain the induced electric field.

\section{Acknowledgments}

\section{Appendix}\label{sec10}

\subsection{Modern GT with low frequency electromagnetic perturbation}\label{sec11}

The guiding center zero order 1-form is
\begin{equation}\label{g2}
{\bar \gamma _0} = \left( {e{{{\bf{\bar A}}}_0} + m\bar U{\bf{\bar b}}} \right)\cdot d{\bf{\bar X}} + \frac{m}{e}\bar \mu d\bar \theta  - (\bar \mu \bar B + \frac{1}{2}{m \bar U^2})dt,
\end{equation}
where the guiding center coordinates plus time is $\mathbf{\bar Z} = ({\bf{\bar X}},\bar {U},\bar {\mu} ,\bar{ \theta} , t)$. And the first order 1-form due to the perturbed magnetic potential vector is
\begin{equation}\label{g3}
\begin{array}{l}
{{\bar \gamma }_1} = e{{{\bf{\bar A}}}_1}({\bf{\bar X}} + {{{\bar{\hat \rho }}}_0},t)\cdot d\left( {{\bf{\bar X}} + {{{\bar{\hat \rho }}}_0}} \right)\\
 \approx e{{{\bf{\bar A}}}_1}\left( {{\bf{\bar X}},t} \right)\cdot d{\bf{\bar X}} + e\left( {{{{\bar{\hat \rho }}}_0}\cdot\nabla } \right){{{\bf{\bar A}}}_1}\left( {{\bf{\bar X}},t} \right)d{\bf{\bar X}}\\
 + e{{{\bf{\bar A}}}_1}\left( {{\bf{\bar X}},t} \right)\cdot \nabla {{{\bar{\hat \rho }}}_0} d{\bf{\bar X}} + e{{{\bf{\bar A}}}_1}\left( {{\bf{\bar X}},t} \right)\cdot\frac{{\partial {{\bar {\hat\rho }}_0}}}{{\partial \bar \mu }}d\bar \mu  + e{{{\bf{\bar A}}}_1}\left( {{\bf{\bar X}},t} \right)\cdot\frac{{\partial {{\bar {\hat \rho} }_0}}}{{\partial \bar{\theta} }}d\bar \theta
\end{array}
\end{equation}
Here, only the first order terms are kept in Eq.(\ref{g3}), $\mathbf{A}_1$ is given in Eq.(\ref{g4}). To get rid of the $\theta$ dynamics in the perturbed 1-form $\bar{\gamma}_1$ in Eq.(\ref{g3}), the coordinate transformation
\begin{equation}\label{g4}
{\bf{Z}} = {e^{{\bar{G}^i}{\partial _{{\bar{Z}^i}}}}}{\bf{\bar Z}}
\end{equation}
is made by transforming $\bar {\bf{Z}}  = (\bar {\bf{X}} ,\bar U,\bar \mu ,\bar \theta ,t) \to {\bf{Z}} = ({\bf{X}},U,\mu ,\theta ,t) $. The center idea of modern GT is to find a group of $\bar{G}_i$ and an auxiliary gauge function $S$ to make all the factors ${\Gamma}_{1i}({\mathbf{Z}})$ in the new first order 1-form ${\Gamma}_1({\mathbf{Z}})$ equal zero except for $\Gamma_{10}={H}_1( \mathbf{Z})$, which is chosen to avoid the secularity of the gauge function $S_1$. Eq.(\ref{g4}) induces a transformation between 1-form like
\begin{equation}\label{g5}
{\Gamma _1}\left( {\bf{Z}} \right) = {\bar \gamma _1}\left( {\bf{Z}} \right) - {L_{G({\bf{Z}})}}{\bar \gamma _0}\left( {\bf{Z}} \right) + d{S_1}\left( {\bf{Z}} \right) - {H_1}({\bf{Z}})dt.
\end{equation}
In this paper, 1-form transformation is carried out up to the first order. By requiring $\Gamma_{1i}=0$ except $i=0$, the equations of $G^i$s are
\begin{equation}\label{a1}
{{{\bf{ G}}}_X} =  - \frac{1}{{eB}}\left( {e{\bf{ b}} \times {{{\bf{ A}}}_1} + {\bf{ b}} \times \nabla {{ S}_1}} \right) - \frac{{\bf{b}}}{m}\frac{{\partial {{ S}_1}}}{{\partial  U}}
\end{equation}
\begin{equation}\label{a2}
{{ G}^U} =  \frac{e}{m}{\bf{ b}}\cdot{{{\bf{ A}}}_1} + \frac{1}{m}{\bf{ b}}\cdot\nabla {{ S}_1}
\end{equation}
\begin{equation}\label{a3}
{{ G}^\mu } = \frac{e}{m}\left( {e{{{\bf{ A}}}_1}\cdot\frac{{\partial {\hat{\rho} _0}}}{{\partial \theta }} + \frac{{\partial {{ S}_1}}}{{\partial \theta }}} \right)
\end{equation}
\begin{equation}\label{a4}
{{ G}^\theta } =  - \frac{{{e^2}}}{m}{{{\bf{ A}}}_1}\cdot\frac{{\partial {\hat{\rho} _0}}}{{\partial  \mu }} - \frac{{\partial {{ S}_1}}}{{\partial  \mu }},
\end{equation}
and the equation for gauge function $S_1$ is
\begin{equation}\label{g6}
\begin{array}{l}
\frac{{\partial {S_1}}}{{\partial t}} + \frac{{eB}}{m}\frac{{\partial {S_1}}}{{\partial \theta }} + U{\bf{b}}\cdot\nabla {S_1} + \frac{\mu }{{eB}}\left( {{\bf{b}} \times \nabla B} \right)\cdot\nabla {S_1} - \frac{\mu }{m}\left( {{\bf{b}}\cdot\nabla B} \right)\frac{{\partial {S_1}}}{{\partial U}} \\
 =  - \frac{{{e^2}B}}{m}{{\bf{A}}_1}\cdot\frac{{\partial {\hat{\rho} _0}}}{{\partial \theta }} - eU{{\bf{A}}_1}\cdot{\bf{b}} - \frac{\mu }{B}\left( {{\bf{b}} \times \nabla B} \right)\cdot{{\bf{A}}_1} - {H_1}.
\end{array}
\end{equation}
The second term on the right hand side is much smaller than the first term. The solution of $S_1$ in Eq.(\ref{g6}) is solved order by order. Firstly the gyroangle averaging is carried out. The symbol $\left\langle {} \right\rangle $ in this paper represents the quantity after gyroangle averaging. If defining ${F_1} = \left\langle { - eU{\bf{b}}\cdot{{\bf{A}}_1} - \frac{\mu }{B}\left( {{\bf{b}} \times \nabla B} \right)\cdot{{\bf{A}}_1}} \right\rangle $, for low frequency electromagnetic perturbation the lowest order equation is
\begin{equation}\label{h1}
\begin{array}{l}
\frac{{eB}}{m}\frac{{\partial {S_{10}}}}{{\partial \theta }} =  - \frac{{{e^2}B}}{m}{{\bf{A}}_1}\left( {{\bf{X}},t} \right)\cdot\frac{{\partial {\hat{\rho} _0}}}{{\partial \theta }}  \\
- eU{\bf{b}}\cdot{{\bf{A}}_1}\left( {{\bf{X}},t} \right) - \frac{\mu }{B}\left( {{\bf{b}} \times \nabla B} \right)\cdot{{\bf{A}}_1} - {F_1}
\end{array}
\end{equation}
which relates the fast variation of $S_1$ to the gyroangle. The next order equation is
\begin{equation}\label{h2}
\begin{array}{l}
\frac{{\partial {S_{11}}}}{{\partial t}} + U{\bf{b}}\cdot\nabla {S_{11}} + \frac{\mu }{{eB}}\left( {{\bf{b}} \times \nabla B} \right)\cdot\nabla {S_{11}} - \frac{\mu }{m}\left( {{\bf{b}}\cdot\nabla B} \right)\frac{{\partial {S_{11}}}}{{\partial U}}  \\
 = {F_1} - {H_1}
\end{array}
\end{equation}

When the resonance happens, $F_1$ is a quantity independent of time, and therefore introduce secularity to $S_{11}$ if integrating Eq.(\ref{h2}) over time. This secularity of $S_{11}$ can be removed by defining $H_1=F_1$ to cancel $F_1$. Eventually, the total 1-form is
\begin{equation}\label{g8}
\Gamma  = \left( {e{{\bf{A}}_0} + mU{\bf{b}}} \right)\cdot d{\bf{X}} + \frac{m}{e}\mu d\theta  - (\mu B + \frac{1}{2}m{U^2} + {H_1})dt,
\end{equation}
by combining the zero order 1-form and the left first order 1-form. The kinetic equations can be derived by applying Euler-Lagrangian equation to the Lagrangian gotten from the 1-form in Eq.(\ref{g8})
\begin{equation}\label{g9}
{\bf{\dot X}} = \left( {\frac{{{\bf{b}} \times \nabla H}}{{eB_\parallel ^*}} + \frac{{{{\bf{B}}^*}}}{{mB_\parallel ^*}}\frac{{\partial H}}{{\partial U}}} \right),
\end{equation}
\begin{equation}\label{g10}
\dot \mu  =  0,
\end{equation}
\begin{equation}\label{g11}
\dot U =  - \frac{{{{\bf{B}}^*}}}{{mB_\parallel ^*}}\cdot\nabla H ,
\end{equation}
\begin{equation}\label{g12}
\dot \theta  = \frac{e}{{{m}}}B_0 ,
\end{equation}
where
\begin{equation}\label{g13}
{{\bf{B}}^*} =  {{\bf{B}} + \frac{m}{e}U\nabla  \times {\bf{b}}}  ,
\end{equation}
\begin{equation}\label{g14}
B_\parallel ^* = {\bf{b}}\cdot{{\bf{B}}^*} ,
\end{equation}
\begin{equation}\label{g15}
H = H_0 + {{\rm{H}}_1}.
\end{equation}
\begin{equation}\label{o1}
H_0=\mu B + \frac{1}{2}m{U^2}
\end{equation}
\begin{equation}\label{n1}
{H_1} = \left\langle { - eU{\bf{b}}\cdot{{\bf{A}}_1}\left( {{\bf{X}},t} \right) - \frac{\mu }{B}\left( {{\bf{b}} \times \nabla B} \right)\cdot{{\bf{A}}_1}} \right\rangle
\end{equation}

\subsection{The modified GT}\label{sec12}

Compared with modern GT which requires $\Gamma_{1i}=0$ for each $i$, our modified edition GT requires $\Gamma_{1i}=0$ for $i={U},\mu,\theta$ and $\Gamma_{1\mathbf{X}}=e\mathbf{A}_1\cdot d\mathbf{X}$ after the operation of LPTT. Carrying out LPTT to the first order, the equations of $G^i$s are
\begin{equation}\label{g16}
{{\bf{G}}_{\bf{X}}} =  - \frac{1}{{eB}}{\bf{b}} \times \nabla {S_1} - \frac{{\bf{b}}}{m}\frac{{\partial {S_1}}}{{\partial U}}
\end{equation}
\begin{equation}\label{g17}
{G^U} = \frac{1}{m}{\bf{b}}\cdot\nabla {S_1}
\end{equation}
\begin{equation}\label{g18}
{G^\mu } = \frac{e}{m}\left( {e{{\bf{A}}_1}\cdot\frac{{\partial {\hat{\rho} _0}}}{{\partial \theta }} + \frac{{\partial {S_1}}}{{\partial \theta }}} \right)
\end{equation}
\begin{equation}\label{g19}
{G^\theta } =  - e{{\bf{A}}_1}\left( {{\bf{X}},t} \right)\cdot\frac{{\partial {\hat{\rho} _0}}}{{\partial \mu }} - \frac{{\partial {S_1}}}{{\partial \mu }}
\end{equation}
The equation of gauge function $S_1$ is
\begin{equation}\label{g20}
\begin{array}{l}
\frac{{\partial {S_1}}}{{\partial t}} + \frac{{eB}}{m}\frac{{\partial {S_1}}}{{\partial \theta }} + U{\bf{b}}\cdot\nabla {S_1} + \frac{\mu }{{eB}}\left( {{\bf{b}} \times \nabla B} \right)\cdot\nabla {S_1} - \frac{\mu }{m}\left( {{\bf{b}}\cdot\nabla B} \right)\frac{{\partial {S_1}}}{{\partial U}}\\
 =  - \frac{{{e^2}B}}{m}{{\bf{A}}_1}\cdot\frac{{\partial {\hat{\rho} _0}}}{{\partial \theta }} - {H_1},
\end{array}
\end{equation}
where $ - \frac{{{e^2}B}}{m}{{\bf{A}}_1}\cdot\frac{{\partial {\rho _0}}}{{\partial \theta }}$ belongs to the lowest order equation of $S_1$. And no term existing in Eq.(\ref{g20}) to introduce secularity to $S_1$.  The first order energy is chosen to be $H_1=0$. And the new total 1-form up to the first order of $O(\varepsilon)$ is
\begin{equation}\label{g21}
\Gamma  = \left( {e{{\bf{A}}_0} + e{{\bf{A}}_1} + mU{\bf{b}}} \right)\cdot d{\bf{X}} + \frac{m}{e}\mu d\theta  - (\mu B + \frac{1}{2}m{U^2})dt.
\end{equation}
By applying Euler-Lagrangian equation to the Lagrangian obtained from the 1-form in Eq.(\ref{g21}), or by the Hamiltonian equations Eq.(18) in Ref.(\cite{1983cary}) for general Hamiltonian system in that paper
\begin{equation}\label{i1}
\frac{{d{z^j}}}{{d{z^0}}} = {J^{jk}}\left( {\frac{{\partial {\gamma _k}}}{{\partial {z^0}}} - \frac{{\partial {\gamma _0}}}{{\partial {z^k}}}} \right)
\end{equation}
where $J^{jk}$ is the Lagrangian Bracket, $\gamma_0$ is Hamiltonian and $\gamma_k$ is the $k$th component of Lagrangina 1-form,
the corresponding kinetic equations are derived as
\begin{equation}\label{g22}
\mathop {\bf{X}}\limits^.  = \frac{{{\bf{b}} \times \nabla {H_0}}}{{eB_\parallel ^*}} + \frac{{U{\bf{B}}{\rm{*}}}}{{B_\parallel ^*}} + \frac{{\partial {{\bf{A}}_1}/\partial t \times {\bf{b}}}}{{B_\parallel ^*}},
\end{equation}
\begin{equation}\label{g23}
\dot \mu  = 0,
\end{equation}
\begin{equation}\label{g24}
\dot U =  - \frac{{{{\bf{B}}^*}}}{{mB_\parallel ^*}}\cdot\nabla \left( {\mu B_0} \right) - \frac{e}{m}{\bf{b}}\cdot\frac{\partial }{{\partial t}}{{\bf{A}}_{\bf{1}}},
\end{equation}
\begin{equation}\label{g25}
\dot \theta  = \frac{e}{m}B_0,
\end{equation}
where ${{\bf{B}}^*} = {{\bf{B}}_0} + {{\bf{B}}_1} + \frac{m}{e}U\nabla  \times {\bf{b}}$.
%

%
%

\newpage
\section*{References}

\bibliographystyle{pst}
\bibliography{magneticperturbation}

\end{document}